\documentclass[prl,amsmath,amssymb,unsortedaddress,superscriptaddress,twocolumn,showpacs]{revtex4-1}

\usepackage[usenames]{color}
\usepackage{amssymb}
\usepackage{grffile}
\usepackage[pdftex]{graphicx}
\usepackage{amsmath, amstext, amssymb, amsfonts, amsxtra}
\usepackage{textcomp}
\usepackage{xspace}
\usepackage{icomma}
\usepackage[normalem]{ulem}

\def \ell{{d}}

\newcommand{\bra}[1]{\ensuremath{\langle #1 \vert}\xspace}%

\newcommand{\ket}[1]{\ensuremath{\vert #1 \rangle}\xspace}%
\newcommand{\aver}[1]{\ensuremath{\langle #1 \rangle}\xspace}%

\newcommand{\bop}{\hat{b}} 
\newcommand{\nop}{\hat{n}} 
\newcommand{\bdop}{\hat{b}^{\dagger}} 
\newcommand{\rhov}{\boldsymbol{\rho}}      
\newcommand{\rhom}{\hat{\rho}}             
\newcommand{\Gf}{{\Gamma}}   

 \newcommand{\sutd}{Singapore University of Technology and Design, 20 Dover Drive, 138682 Singapore.}
 \newcommand{\unige}{D\'epartement de Physique Th\'eorique, Universit\'e de Gen\`eve, CH-1211 Gen\`eve, Switzerland.}
 \newcommand{\ubc}{Department of Physics and Astronomy, University of British Columbia, Vancouver V6T 1Z1, Canada.} 
 \newcommand{\cf}{Coll\`ege de France, 11 place Marcelin Berthelot, 75005 Paris, France.} 
 \newcommand{\ep}{Centre de Physique Th\'eorique, Ecole Polytechnique, CNRS, 91128 Palaiseau Cedex, France.}
 \newcommand{\dpmc}{DPMC-MaNEP,  Universit\'e de Gen\`eve, CH-1211 Gen\`eve, Switzerland.}

\begin{document}

\title{Interaction-induced impeding of decoherence and anomalous diffusion}             

  \author{Dario Poletti}
  \affiliation{\unige}
  \affiliation{\sutd}
  \author{Jean-S\'ebastien Bernier}
  \affiliation{\ubc} 
  \author{Antoine Georges}
  \affiliation{\cf}
  \affiliation{\ep}
  \affiliation{\dpmc}
  \author{Corinna Kollath}
  \affiliation{\unige}   
  \affiliation{\ep}

\begin{abstract}
We study how the interplay of dissipation and interactions affects the dynamics of a bosonic many-body quantum system. 
In the presence of both dissipation and strongly repulsive interactions, observables such as the coherence 
and the density fluctuations display three dynamical regimes: 
an initial exponential variation followed by a power-law regime, and finally a slow exponential 
convergence to 
their asymptotic values. 
These very long-time scales arise as dissipation forces the population of states disfavored by
interactions. 
The long-time, strong coupling dynamics are understood by performing a mapping onto a 
classical diffusion process 
displaying non-Brownian behavior. 
While both dissipation and strong interactions tend to suppress coherence when acting separately, 
we find that strong interaction impedes the decoherence process generated by the dissipation. 
\end{abstract}

\pacs{03.75.Gg, 67.85.De, 03.65.Yz, 67.85.Hj} 

\maketitle

Understanding the influence of the environment on the dynamics of physical systems 
is of paramount importance in the development of quantum-based technologies.
For systems as diverse as those encountered in solid-state 
physics~\cite{LeggettZwerger1987,SchoenZaikin1990,BreuerPetruccione2002,Weissbook2008}, atomic and 
molecular physics~\cite{CohenTannoudjiGrynberg1998} 
or quantum optics~\cite{GardinerZollerBook,CarmichaelBook}, 
coupling to an environment often results in the loss of coherence.
Devising methods to control and suppress decoherence is therefore at the
heart of various present and future applications such as 
high-precision clocks, 
measuring devices~\cite{CroninPritchard2009,BudkerRomalis2007},
and quantum computers~\cite{NielsenChuangBook}. 
A clear understanding of the effects of an environment on correlated quantum many-body systems is presently lacking. 
For example, the consequences resulting from the interplay of interactions and dissipation are not well known.
Even though each process individually tends to suppress coherence, it is not understood if they
cooperate or destructively compete when acting simultaneously. 
An ideal testbed to answer such questions is provided by cold atomic gases as, in these systems, both
the interactions and dissipative processes are highly controllable.

Exciting phenomena have already been proposed to occur when cold gases are coupled to an environment: 
a Zeno-like behavior due to local atom losses in systems of interacting atoms
confined to double wells~\cite{KhodorkovskyVardi2008,ShchesnovishKonotop2010,ShchesnovichMogilevtsev2010} 
or optical lattices~\cite{BarmettlerKollath2011}; 
improved stability against weak dissipative effects for bosonic atoms in both the weakly interacting~\cite{WitthautWimberger2011} 
and Mott-insulating regimes~\cite{PichlerZoller2010}; a dynamical phase transition between a condensed and 
a thermal steady state~\cite{TomadinZoller2011} for a coherence-enhancing dissipative process; and
the possibility to engineer dark states with highly desirable
properties~\cite{SyassenDuerr2008,DiehlZoller2008,Garcia-RipollCirac2009,TomadinZoller2011,KantianDaley2009,VerstraeteCirac2009}.

In this work, we investigate the influence of a Markovian (memory-less) environment, e.g.~a light field or a thermal cloud, 
on strongly interacting bosonic atoms in a double well potential. We find that dissipation and interactions destructively compete, 
resulting in two surprising phenomena:     
(i) contrary to naive intuition, decoherence is slowed down in the presence of strong interactions,
(ii) at intermediate times, physical quantities such as the coherence and density fluctuations are found to display a
 slow power-law time-dependence 
characterized by a time-scale $t^*$ which is considerably enhanced by interactions. 

The algebraic regime can be characterized in detail in the limit of a large number of atoms $N$ by mapping the evolution 
onto a classical diffusion in the configuration space of all Fock states. 
For $N$ bosons confined to two wells, this space is described by a 
single coordinate $x$. The diffusion rate is found to be strongly 
dependent on the configuration: it is large (small) for configurations which have 
a low (high) cost in interaction energy. As shown analytically, this translates into a  
non-Brownian diffusion~\cite{BouchaudGeorges1990} with $\sqrt{\langle x^2 \rangle} \sim \tau^{1/4}$ and a 
non-Gaussian probability density 
$p(x,\tau) \propto \tau^{-1/4}\,e^{-x^4/4\tau}$. Here $\tau=t/t^*$, with the long time-scale $t^*$ given by 
$\gamma t^*=2N^2 (U/J)^2$ ($\gamma$ is the Markovian dissipation rate, 
$U$ the interaction strength and $J$ the tunneling coefficient between the two wells).  

Considering $N$ bosonic atoms trapped in a sufficiently deep optical lattice with $L$ sites, 
the evolution of the system is described by the master equation 
\begin{equation}
 i\hbar\partial_t\hat{\rho}=[\hat{H},\hat{\rho}]+i\hbar\;\mathcal{D}\left(\hat{\rho}\right). \label{eq:master}     
\end{equation}
The first term $[\hat{H},\hat{\rho}]$ describes the unitary evolution of the density 
matrix $\hat{\rho}$ governed by the Hamiltonian   
$\hat{H}=-J\sum_{l=1}^{L-1} \left(\bdop_l\bop_{l+1}+h.c.\right)+\frac{U}2 \sum_{l=1}^{L} \nop_l\left(\nop_l-1\right)$.
The operators $\bdop_l$ and $\bop_l$ are bosonic creation and annihilation operators on site $l$ and $\nop_l=\bdop_l\bop_l$ 
counts the number of atoms. 
At large $U/J$ the atoms tend to be more localized which typically results in a reduction of 
the coherence between neighboring sites
$C=\sum_{l} \langle\bdop_l\bop_{l+1} + \bdop_{l+1}\bop_l\rangle$.
The dissipative process is modeled by  
$\mathcal{D}\left(\hat{\rho}\right)
=\gamma \sum_{l=1}^L\left(\nop_l\hat{\rho}\nop_l-\frac 1 2 \nop_l^2\hat{\rho}-\frac 1 2 \hat{\rho}\nop_l^2\right)$, 
where $\gamma$ is the coupling to the environment and the quantum jump operators correspond to local density operators $\hat{n}_l$. 
This form of dissipation has been identified as one of the most important heating processes 
in the presence of a red-detuned optical lattice potential~\cite{GerbierCastin2010,PichlerZoller2010}. 
In the absence of interactions ($U=0$), it causes a rapid exponential loss of coherence $C(t)= C(0) e^{-\gamma t}$.   

It is useful to recast the quantum master 
equation (\ref{eq:master}) into an eigenvalue equation $M\rhov=- \lambda \rhov$, with $\rhov_{\alpha}$ and $-\lambda_{\alpha}$ being respectively eigenvectors and eigenvalues of the matrix $M$. 
The positive real parts of $\lambda_\alpha$ are the inverse relaxation times, since 
 $\rhov_{\alpha}(t)=e^{-\lambda_\alpha t}\rhov_\alpha(0)$. 
Hence, the longest living states are associated with the smallest $\mathrm{Re}(\lambda_\alpha)$. 
For $J=0$, all combinations of diagonal Fock states $|\{n_l\}\rangle\langle \{n_l\} |$ are steady-states ($\lambda_{\alpha}=0$). 
However, for $J\not = 0$ (and $L$ finite), the unique steady-state is the 
completely mixed state  $\rhom_{\!_S}\propto \sum_{\{n_l\}}|\{n_l\}\rangle\langle \{n_l\} |$ in which all Fock states 
have equal weight (corresponding to maximal Von-Neumann entropy). We will show that $\rhom_{\!_S}$ is reached in a highly non-trivial way. 

\begin{figure}[!ht]
 \centering\includegraphics[width=0.9\columnwidth]{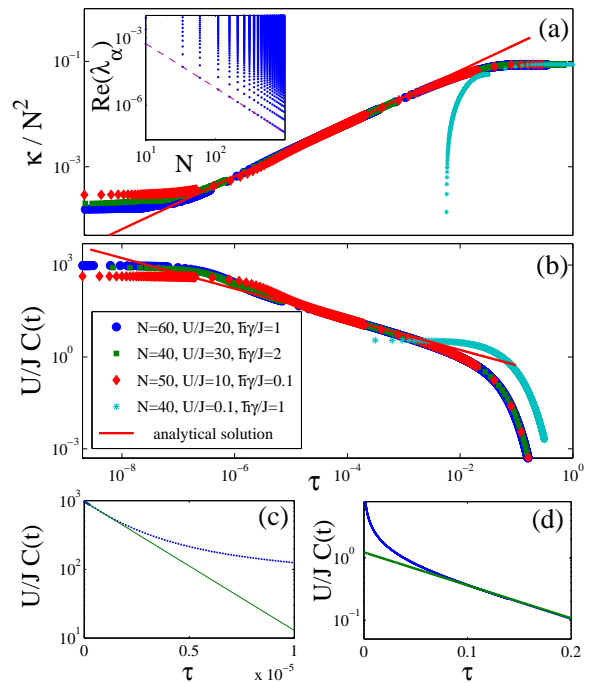}
   \caption{(color online) 
(a-b) Density fluctuations (a) and rescaled-coherence (b) versus rescaled time $\tau$. A data collapse is observed for 
large interaction strength $U$ and atom number $N$, independently of the other system parameters. 
The analytical results are taken from Eq.(\ref{eq:lity}) and (\ref{eq:coherence}). 
The inset of (a) shows the real parts of the smallest $\lambda_{\alpha}$ versus $N$ and a fit $\propto 1/N^{2}$.    
(c-d) Coherence decay at short (c) and long (d) times ($N=60$, $U/J=20$, and $\hbar\gamma/J=1$) with 
exponential fit (continuous green line). $L=2$. 
} \label{fig:Fig1} 
\end{figure}

To pinpoint this, we concentrate on the two site 
Bose-Hubbard model ($L=2$). Various aspects of the influence of dissipation have been studied in this set-up, 
focusing mainly on weak interactions~\cite{LouisSavage2001,HuangMoore2006,DalvitVishik2006,Bar-GillKurizki2011,FerriniHekking2010}. 
For our study the atoms are prepared in the symmetric ground state of the Hamiltonian and the dissipation 
is turned on at time $t=0$ \cite{Note2}. 
We monitor different experimentally measurable quantities, such as the coherence between the two wells $C$, the 
local density fluctuations $\kappa=\aver{n_1^2}-\aver{n_1}^2$, 
and the probability $P_b$ to measure a balanced configuration with $N/2$ particles in each well. 
At low interaction strength $U$ or small particle number $N$, exponential evolutions dominate. Their time-scales depend 
crucially on the parameter regime considered. 
In contrast, for sufficiently large $N$ and $UN\gg\hbar\gamma$, three distinct regimes can be identified: 
(i) a fast exponential variation at short times ($\gamma t\ll 1$), (ii) a drastic slowing down associated with a power-law regime 
at intermediate times ($1/\gamma\ll t\ll t^*$), and (iii) a slow exponential approach to the asymptotic value 
at long times ($t\gtrsim t^*$). These regimes are exemplified in Fig.~\ref{fig:Fig1} for the rise of 
density fluctuations and the decay of coherence obtained by solving Eq.~(\ref{eq:master}) numerically.  

A first insight into the origin of regimes (ii) and (iii) can be  gained by analyzing the 
eigenvalue spectrum (inset of Fig.~\ref{fig:Fig1}(a)).  
The final exponential regime is due to the fact that the lowest non-vanishing real part of an 
eigenvalue $\lambda \propto \gamma J^2/U^2N^2$ is isolated. 
The algebraic regime arises from the complex interplay of many slowly decaying states $\rhov_{\alpha}$ 
(having non-zero overlap with the initial 
state)
corresponding to the band of eigenvalues in the inset of Fig.~\ref{fig:Fig1}(a). 

To understand quantitatively the different dynamical behaviors, we reduce the system to a simpler model for $t\gg 1/\gamma$ and strong interaction $U\gg J,\;\hbar\gamma$. This in turn can be mapped using the large $N$ limit onto a {\it classical} diffusion problem in configuration space. 
We represent the density matrix in the basis of Fock states as $ \hat{\rho}=\sum \rho_{n,m} \ket{n}\bra{m}$ where $n,m=0,1,\dots, N$ 
labels the number of atoms in the left well.
In 
this regime, it is justified to use adiabatic elimination 
and map the full evolution onto an equation for the diagonal elements $\rho_{n,n}$ only. 
This is done by integrating $\partial_t \rho_{n,n+1}$ (derived from Eq.(\ref{eq:master})) by noticing that the diagonal terms of the 
density matrix are slowly varying compared to the off-diagonal ones (following \cite{SaitoKayanuma2002} and for even $N$). 
This gives 
\begin{eqnarray} 
    \rho_{n,n+1}(t)&\approx&J_R \left(1-\frac{i\hbar \gamma}{(N-2n-1)U}\right)\Delta \rho_{n}(t) \label{eq:offdiagevo}  
\end{eqnarray} 
where $J_R=\frac{J\sqrt{W_{n+1}}}{2UN^2}$,
$W_{n+1}= \frac{(n+1)(N-n)}{(n-N/2+1/2)^2}$, and $\Delta \rho_{n}=N^2(\rho_{n+1,n+1}-\rho_{n,n})$.  
Using (\ref{eq:offdiagevo}) into $\partial_t \rho_{n,n}$, one obtains
\begin{eqnarray} 
 \partial_\tau \rho_{n,n}&=& 2\left(W_{n+1}\Delta\rho_{n} - W_n \Delta\rho_{n-1}\right). 
\label{eq:mastersimp}
\end{eqnarray} 
Remarkably, the interaction, dissipative coupling, tunneling amplitude and particle number only enter the equations 
for the diagonal elements $\rho_{n,n}$ via the dimensionless time $\tau=t/t^*$. 
The emerging time-scale $t^*=\frac{2N^2 U^2}{\gamma J^2}$ controls the long-time behavior of the system,
and becomes very large 
for strong coupling $U$ and large $N$ \cite{Note}. 
Additionally, this dependence explains the scaling of the lowest eigenvalues.

 
We now perform the large $N$ limit 
calculation
in which the discrete master equation~(\ref{eq:mastersimp}) is mapped onto 
a classical diffusion equation in the continuum. 
We map the configuration space onto a coordinate space by $x=n/N-1/2\in [-1/2,1/2]$ which becomes continuous in the large $N$ limit. 
The boundaries ($x\approx \pm 1/2$) correspond to strongly imbalanced occupation of the double well, 
whereas the center ($x=0$) corresponds to the balanced configuration.
The diagonal elements of the density matrix are related to a probability density $p(x,\tau)$  
by $N\rho_{n,n}(\tau)=p(x,\tau)$, with normalization $\int_{-1/2}^{1/2}p(x)\textrm{d} x =1$ 
(insuring $\mathrm{tr}\,\hat{\rho}=1$). 
We thus obtain the diffusion equation
\begin{eqnarray}
\partial_x\left[D(x)\partial_x p(x,\tau)\right]=\partial_\tau \; p(x,\tau).  \label{eq:continuum}
\end{eqnarray}   
The diffusion function $D(x)=\frac{1}{4x^2}-1$ is strongly dependent on the variable $x$: 
it diverges at the center ($x=0$) and vanishes at the boundaries ($x\approx\pm 1/2$). 
Physically this slow diffusion at the boundaries corresponds to the slow population of the energetically costly imbalanced configurations. 
  
\begin{figure}[!ht]
 \includegraphics[width=\columnwidth]{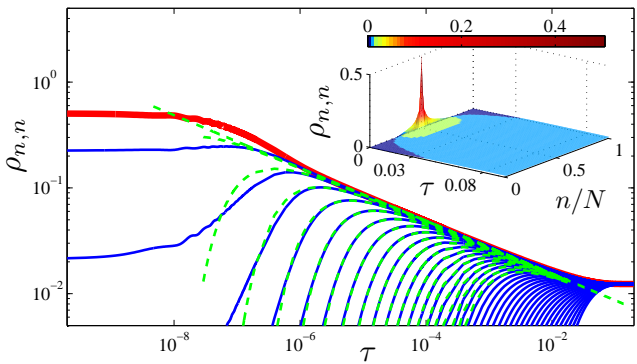}
  \caption{(color online) Diagonal terms of the density matrix $\rho_{n,n}$  versus rescaled time $\tau$ in 
log-log form: the exact solutions of Eq.~(\ref{eq:master}) for $n=41$(top)$...80$(bottom)
are represented by blue continuous lines; the element $n=40$ (thick red line) is proportional to the probability of finding 
a balanced configuration $P_b$; the corresponding diffusion density $p(x=n/N,\tau)$ Eq.~(\ref{eq:proba}) up 
to $n/N=60$ are shown in dashed green lines. Inset: 3D plot of the evolution. 
Parameters: $U/J=20$, $\hbar\gamma/J=1$, $N=80$ and $L=2$.} \label{fig:Fig2}      
\end{figure}

Within this mapping, the initial ground state corresponds to $p(x,\tau=0)$ and is peaked at $x=0$.
The asymptotic long time limit is the uniform distribution $p(x,\tau=\infty)=1$ representing the totally mixed state $\rhom_S$. 
To gain more insight into the actual diffusion process, we insert the scaling 
ansatz $p(x,\tau)= \frac{1}{\tau^{\nu}}f(\xi)$ with $\xi=x/\tau^{\nu}$ 
into (\ref{eq:continuum}). We find that a scaling solution exists provided $\nu=1/4$ and $\tau\ll 1$. 
The function $f(\xi)$ satisfies $\xi f''+(\xi^4-2)f'+\xi^3f=0$, whose solution can be found in closed form: 
$f(\xi)\propto \exp(-\xi^4/4)$. Therefore, the diffusion process at short rescaled 
time ($\tau\ll 1$, i.e. $t\ll t^*$) is 
non-Brownian and described by
\begin{equation}
p(x,\tau)= \frac{\sqrt{2}}{\Gf(1/4)} \frac{1}{\tau^{1/4}}\,\exp\left(-x^4/4\tau\right) \label{eq:proba}    
\end{equation}
where $\Gamma(1/4)$ in the normalization denotes the gamma function.    
Compared to normal diffusion given by $p(x,\tau)= 1/\sqrt{4\pi\sigma\tau} \exp(-x^2/4\sigma\tau)$ for a space-{\it independent} 
diffusion constant $\sigma$, the divergence in $D(x)$ for $x\approx 0$ leads to a highly accelerated initial diffusion.  

The analytical expression~(\ref{eq:proba}) reproduces accurately the exact evolution of the diagonal terms of the 
density matrix computed from~(\ref{eq:master}) as shown in Fig.~\ref{fig:Fig2}.
The very fast initial broadening of the peak is followed by a much slower occupation of the boundary (imbalanced) regions. 
In particular, we observe a clear power-law decay (with exponent $1/4$) in the 
exact evolution of the density matrix elements, in very good agreement 
with the analytical form~(\ref{eq:proba}) even for
a relatively low number of atoms. For times $\tau=t/t^*\sim 0.1$ (time-scales which can actually 
be very long in an experiment), the lowest eigenvalue 
of the discrete spectrum of~(\ref{eq:continuum}) due to the finite atom number controls the eventual 
exponential convergence of $p(x,\tau)$ to the uniform distribution. 

\begin{figure}[!ht] 
  \centering 
  \includegraphics[width=0.9\columnwidth]{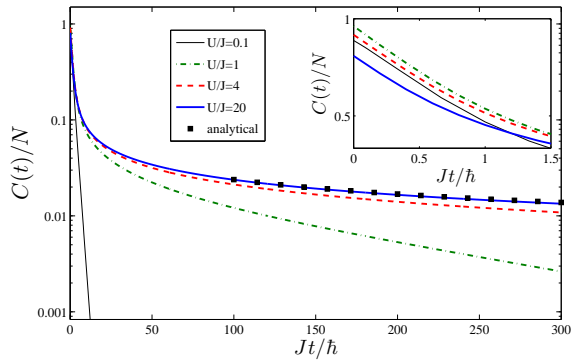}
  \caption{(color online) Coherence divided by number of atoms versus time for different interactions 
and for $N=60$, $\hbar\gamma/J=1$ and $L=2$. Inset: zoom on the short times.  } \label{fig:Fig4}         
 \end{figure}

The diffusion in configuration space has important consequences for experimentally measurable quantities. 
The probability of measuring a balanced configuration $P_b(t)$ 
is directly proportional to 
the central density $p(x=0,t)$. Therefore, its detection would reveal the passage from the initially exponential decay to the algebraic one in 
$\tau^{-1/4}$ (see the red thick line in Fig.~\ref{fig:Fig2}). 
The density fluctuations $\kappa$ show an increase which in the power-law region is well described, using 
$\kappa/N^2=\int_{-1/2}^{1/2}(x^2+x)p(x,\tau)dx$, by     
\begin{equation}
\frac{\kappa}{N^2}=\frac{2\Gf(3/4)}{\Gf(1/4)} \sqrt{\tau}=\frac{\sqrt{2}\Gf(3/4)}{\Gf(1/4)} \frac{J}{NU}\sqrt{t\gamma} . \label{eq:lity}  
\end{equation}
As shown in Fig.~\ref{fig:Fig1}(a), this equation describes accurately the exact density fluctuations at intermediate
 times without any adjustable parameter. 
At longer times, the density fluctuations tends towards its asymptotic value $\frac{\kappa}{N^2}=\frac{1}{12} +\frac{1}{6N}$.  
For the coherence, from the continuum limit of~(\ref{eq:offdiagevo}), 
we can derive a scaling behavior $C(t)=J/U\;{\cal C}(t/t^*)$ 
where ${\cal{C}}(\tau)\approx\int_{-1/2}^{1/2}\frac{x^2-1/4}{x} \partial_x p(x,\tau) {\textrm dx}$. 
 This scaling is verified over a large range of different parameters $U>\hbar\gamma\,,\,J$ as shown in Fig.~\ref{fig:Fig1}(b). 
In the power-law regime, where $\tau$ is small, this reads  
\begin{eqnarray} 
\frac{C(t)}{N}=\frac{\Gf(3/4)}{2\Gf(1/4)} \frac{J}{UN} \frac{1}{\sqrt{\tau}}=\frac{\Gf(3/4)}{\sqrt{2}\Gf(1/4)}  
\frac{1}{\sqrt{\gamma t}} \label{eq:coherence}   
\end{eqnarray}  
with again excellent agreement with the exact results over a large range of intermediate times 
without any adjustable parameter (Fig.~\ref{fig:Fig1}(b) and Fig.~\ref{fig:Fig4}). We note that in this regime 
the coherence only depends on the original dissipative coupling $\gamma$ and not on $J$ or $U$. 

In Fig.~\ref{fig:Fig4} we compare the decay of the coherence for stronger and weaker interaction strengths.  
A weaker interaction strength typically leads to a larger value of the coherence in the initial state and at small times (inset). 
After the initial exponential decay, the regime of algebraic decay $\propto 1/\sqrt{t}$ is 
manifest for large interaction strength $UN \gg \hbar\gamma,~J$. 
This regime shrinks when lowering the interaction, and is followed by the eventual exponential 
decay at long times. 
Surprisingly, the coherence for larger interactions {\it exceeds} that for lower interactions and survives for much longer times.
Hence the presence of interactions impedes the dissipation-induced decoherence.
The physical reason behind this finding is that it is difficult to populate Fock space configurations
which are energetically disfavored by strong interactions. We stress that the slowing 
down of decoherence, unlike the use of particular squeezed, ``decoherence-free'', states \cite{LerouxVuletic2010}, 
generally holds independently of the initial condition. 

To summarize, we found that the time-evolution of the coherence and density fluctuations of strongly 
interacting bosonic systems subject to dissipative effects can present three consecutive regimes: an 
initial fast exponential one, and an intermediate 
power-law regime associated with anomalous diffusion which is eventually followed by a final exponential regime. 
The latter sets in after a time $\sim t^*$ which becomes very large for strong interactions and large number of particles, and 
the decay rate is reduced accordingly. 
Simulations on three-site systems and mean-field calculations for an infinite lattice confirm the existence of the slowing down of decoherence and the presence of a power law regime. 
We understand that a power-law dynamics, atypical for the considered Markovian environment, also occurs for different kinds of Markovian dissipative mechanisms \cite{MarinoSilva2012}. 
%

To probe these phenomena experimentally, one could use a single double well potential or an array 
of decoupled double wells generated by an optical super-lattice potential.
Observing the different regimes proposed here requires strong interactions
while maintaining the single band approximation during the whole experiment.
To meet this requirement, we suggest the use of a light species, for example Lithium, 
whose interaction can be tuned by a Feshbach resonance~\cite{StreckerHulet2002, KhaykovichSalomon2002}. 
The light mass enables the use of a deep optical lattice potential (a band gap of about $250$ kHz is realistically 
achievable) while keeping reasonable time-scales. An alternative realization would be to trap two bosonic 
species in one well and to tune their interaction with a Feshbach resonance~\cite{GrossOberthaler2010}.  

We thank P. Barmettler, H.P. Breuer, J. Dalibard, J.-P. Eckmann, M. Greiner, M. Lukin and V. Vuletic for fruitful discussions. 
We acknowledge ANR (FAMOUS), SNSF (Division II, MaNEP), CIFAR, NSERC of Canada and the DARPA-OLE program for financial support. 

\bibliographystyle{prsty}

\begin{thebibliography}{10}
\bibitem{Note}
The effective reduction of the dissipative coupling from $\gamma $ to $\gamma
  (J/U)^2$ can be understood from a second-order perturbation theory argument
  ($\propto \protect \mathrm {Im}\protect \tmspace +\thinmuskip
  {.1667em}J^2/(U-i\hbar\gamma )$ for $U\gg \hbar\gamma $) analogous to super-exchange.
\end{thebibliography}

\end{document}